\documentclass[amsmath,twocolumn,aps,prl,superscriptaddress]{revtex4-1}

\usepackage{bm}
\usepackage{graphics}
\usepackage{graphicx}
\usepackage{color}
\title{}

\begin{document}

\title{Mobile Dopants in Inhomogeneous Superconductors}
\author{I.V. Sokolovskii}
\affiliation{The First Electrotechnical University «LETI», 197376, St. Petersburg, Russia}
\affiliation{A.F. Ioffe Physico-Technical Institute of
Russian Academy of Sciences, 194021 St. Petersburg, Russia}

\author{S.A. Ktitorov}
\affiliation{A.F. Ioffe Physico-Technical Institute of
Russian Academy of Sciences, 194021 St. Petersburg, Russia}

\author{A.Yu. Zyuzin}
\affiliation{A.F. Ioffe Physico-Technical Institute of
Russian Academy of Sciences, 194021 St. Petersburg, Russia}


\begin{abstract}
We consider a superconductor hosting mobile impurities, which locally change the superconducting transition temperature. The BCS interaction at the impurity is different both in magnitude and in sign from the BCS interaction in the bulk. It is shown that due to the attraction between impurities, they tend to form more condensed state. We also consider the distribution of mobile impurities with local BCS attraction or repulsion at the vicinity of the superconductor-normal metal interface.
\end{abstract}
\maketitle

\section{Introduction}

Since the usual superconductivity is a low-temperature phenomenon 
thermal jumps of point impurities or atoms are practically forbidden in the superconducting state. The coherent propagation at low temperature is possible only for very light
particles, such as muons \cite{bib:Kadono1} and hydrogen \cite
{bib:Hydro-metal}, see also review \cite{bib:rev-RMP}.

The effect of superconductivity on the coherent motion of impurities usually
reduces to a change in the electronic polaron effect \cite{bib:Kagan} due to
a strong modification in the electronic spectrum as a result of the phase transition into
the superconducting state. In particular, the opening of the superconducting gap decreases
the number of electrons at the Fermi level hence increases the coherent
mobility. This is in agreement with the experiment \cite{bib:Kadono2}.

Nowadays the superconducting transition temperatures reaches values of the order of hundred Kelvin. In this case, one might assume 
that impurities are relatively
fast moving by making activation jumps, thus forming a mobile component in a
superconductor. It is a natural question to ask what physical consequences of the presence of
a mobile component can be expected in superconductors and superconducting
structures.

Here we show that if the BCS interaction on impurities is different from its mean
value, then it is energetically favourable to form an inhomogeneous
superconducting state. This result applies to the situations, where the mobile component has local BCS interaction constant, which can be both larger or
smaller as compared to its average value, or even has repultion sign. We also consider several examples of energy profile for mobile component in the superconductor-normal metal structures.

\section{Optimal concentration of mobile impurities}

Naturally, the presence of the mobile component locally modifies the physical properties of superconductor.
Suppose here that this is the superconducting
transition temperature. One might ask, whether the back action of the local change of the critical temperature 
will affect the spatial structure of the mobile component.

To proceed, let us consider the condensation energy of a superconductor, which has the
volume $V$. Within the mean field BCS model and in the situation where the temperature is smaller than
the energy gap, the condensation energy is given by \cite{bib:Sup-book} 
\begin{equation}
E_{\mathrm{cond}}=-\frac{V\nu}{2}|\Delta|^{2}\equiv -\varepsilon _{c}V e^{-2/\lambda\nu},  \label{condens}
\end{equation}
where $\nu$, $\lambda$, and $|\Delta|$ are the density of electronic states at the
Fermi level, BCS electron-electron interaction constant, and superconducting gap,  respectively. The second equality in Eq. \ref{condens} is valid at small temperature.
We note that the exponent in Eq. \ref{condens} is the most sensitive to the material parameters, compared to the energy $
\varepsilon_{c}>0$, which is of the order of Debye frequency.

\begin{figure}
\includegraphics [width = 7cm] {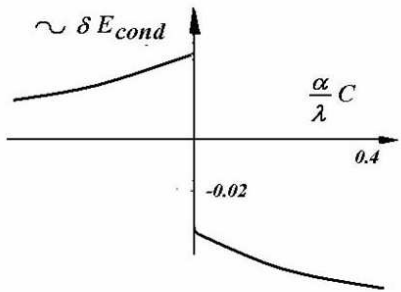}
\caption{The gain of condensation energy as a function of concentration of
injected particles, which increase BCS attraction interaction ($
\alpha C/\lambda >0$) or decrease it ($\alpha C/\lambda <0 $), here $\lambda \nu =0.4 $.} \label{fig1}
\end{figure}

Let us consider $N$ particles injected into the volume $V_{0} \ll V$ of the superconductor. We assume that
they modify the exponent in (\ref{condens}) by increasing or decreasing attraction between the electrons
\begin{equation}
\lambda \rightarrow \lambda + \alpha C,
\end{equation}
where $\alpha$ is some constant.
Here we made an assumption that the variation of the interaction constant is
proportional to the concentration $C=N/V_{0}$ of injected particles. 
The condensation energy of superconductor in the presence of injected particles is given by
\begin{equation}
E_{\mathrm{cond}}= -\varepsilon_{c}[(V-V_{0})e^{-2/\lambda \nu}+V_{0}e^{-2/(\lambda +\alpha C) \nu}],
\end{equation}
where we neglect the surface energy. Hence, the change in the condensation energy after the injection of $N$ particles reads
\begin{eqnarray}\label{energy-concentr}
 \delta E_{\mathrm{cond}}= -\varepsilon_{c}V_{0}[e^{-2/(\lambda +\alpha C)\nu}- e^{-2/\lambda\nu}]\nonumber\\
 \sim -C^{-1}[e^{-2/(\lambda +\alpha C )\nu} - e^{-2/\lambda\nu}].
\end{eqnarray}
The schematic dependence of $ \delta E_{\mathrm{cond}}$ in (\ref{energy-concentr}) on the concentration of injected particles is shown in
Fig. \ref{fig1}.

Let us consider two opposite situations, where the impurities increase 
($\alpha C/\lambda >0$) or decrease ($0>\alpha C/\lambda $) the BCS constant, which has a finite value in the limit $C\rightarrow 0$ in both cases. 
Therefore, a shift of the condensation energy is proportional to number of injected particles. We find gain (loss) of condensation energy in the
situation where the defects increase (decrease) the BCS constant.

In both cases the system gains more condensation energy with the
increase of concentration, which also means that the injected particles prefer more condensed state. Namely, at
a given number of particles the increase of concentration means the decrease of the
volume of injected particles.

The fact of attraction between the injected particles, which locally shift the
superconducting temperature, can be verified within the Ginzburg-Landau
functional approach. Note that the model of a superconductor hosting granules 
with repulsive BCS interaction was considered in \cite{bib:Hrushka}.

Our approach can not be applied for the systems with large concentration $|\alpha C| \approx |\lambda|$, in which expressions in Eq. \ref{energy-concentr} should be corrected. Although, due to the
denominator in (\ref{energy-concentr}) there is always a trend $\delta E_{\mathrm{cond}}\rightarrow 0$ with increasing concentration.
This means the existence of optimal concentration for impurities, which increase attraction. Of course, there are also elastic forces present in real solids. Naturally, the ground state of
mobile particles will be determined by the balance of all forces.

\section{Impurity energy profile near superconductor - normal metal interface}

It is interesting to study the distribution of mobile impurities near the interface of the
superconductor and normal metal. Such distribution is determined by the energy profile for the
mobile defects. To consider the influence of superconductivity on the energy
profile we write the BCS Hamiltonian with spatially dependent interaction
constant, which reflects the fact that near a defect the phonon
spectrum and Coulomb electron-electron interaction might be modified
\begin{equation}
H_{\mathrm{BCS}}=\int d\mathbf{r}\lambda (\mathbf{r})\Psi _{\uparrow }^{+}(\mathbf{r}
)\Psi _{\downarrow }^{+}(\mathbf{r})\Psi _{\downarrow }(\mathbf{r})\Psi
_{\uparrow }(\mathbf{r}).
\end{equation}
Let $\lambda _{0}$ be the bulk BCS interaction constant, such that near the
impurity one has $\delta \lambda (\mathbf{r})=\lambda (\mathbf{r})-\lambda _{0}\neq 0
$. This inequality determines the BCS radius of impurity.

Within the mean field approximation a correction to the energy due to single spinless impurity (we will comment on the spin-flip scattering on magnetic impurity later) is given by
\begin{equation}
\delta \Omega =\int d\mathbf{r}\delta \lambda (\mathbf{r}) 
\langle \Psi _{\uparrow }^{+}(\mathbf{r})\Psi _{\downarrow }^{+}(\mathbf{r}) \rangle
\langle \Psi _{\downarrow }(\mathbf{r})\Psi _{\uparrow }(\mathbf{r})\rangle.
\end{equation}

We consider impurity placed at point $\mathbf{R}$ and assume for simplicity BCS radius of impurity to be of the order of electron wavelength. 
We obtain the position dependent energy
\begin{eqnarray}\label{BCS-energy}\nonumber
\delta\Omega (\mathbf{R}) &=& \int d\mathbf{r} \delta\lambda(\mathbf{r}) |\Delta(
\mathbf{R})|^{2}\\
&=&
\int d\mathbf{r} \delta\lambda(\mathbf{r}
) |T\sum_{\omega_{n}}F(\mathbf{R},\mathbf{R};\omega_{n})|^{2},
\end{eqnarray}
where $|\Delta (\mathbf{R})|$ is the modulus of local value of the Cooper pair
wave-function, $F(\mathbf{R},\mathbf{R^{\prime }};\omega _{n})$ is the anomalous
superconducting Green function, and $\omega _{n}=(2n+1)\pi T$ is Matsubara
frequency.

We now take into account the interaction of BCS impurity with the interface between the superconductor and
normal metal. In normal region the superconducting correlations arise due to the
proximity effect. For temperature $T \sim |\Delta (T)|$ and distance from the interface $%
R\leq v_F/T$, where $v_F$ is the Fermi velocity, we estimate the position dependent energy as
\begin{equation}\label{energy1}
\delta\Omega(\mathbf{R})\sim \left\{ \nu \Delta(T)\ln|RT/v_{F}| \right\}^{2}\int d\mathbf{r%
} \delta\lambda(\mathbf{r}).
\end{equation}
At larger distances, such that $R > v_F/T$, the energy decreases exponentially with the increase of
distance. Noting that if $\int d\mathbf{r} \delta\lambda(\mathbf{r})\sim
\epsilon_{F}/p^{3}_{F}$, where $p_F$ and $\epsilon_F$ are the Fermi momentum and energy respectively, we obtain
\begin{equation}\label{energy2}
\delta\Omega (\mathbf{R})\sim \Delta^{2}(T)/\epsilon_{F}.
\end{equation}

Let us now analyze possible realizations of the energy profiles. We assume
that $\lambda _{0}$ is zero in the normal metal and takes finite
values in the superconductor describing finite attraction between electrons. Therefore, the
factor $\int d\mathbf{r}\delta \lambda (\mathbf{r})$ itself depends on $R$
near the superconductor- normal metal interface.

\begin{figure}
\includegraphics[width = 4cm] {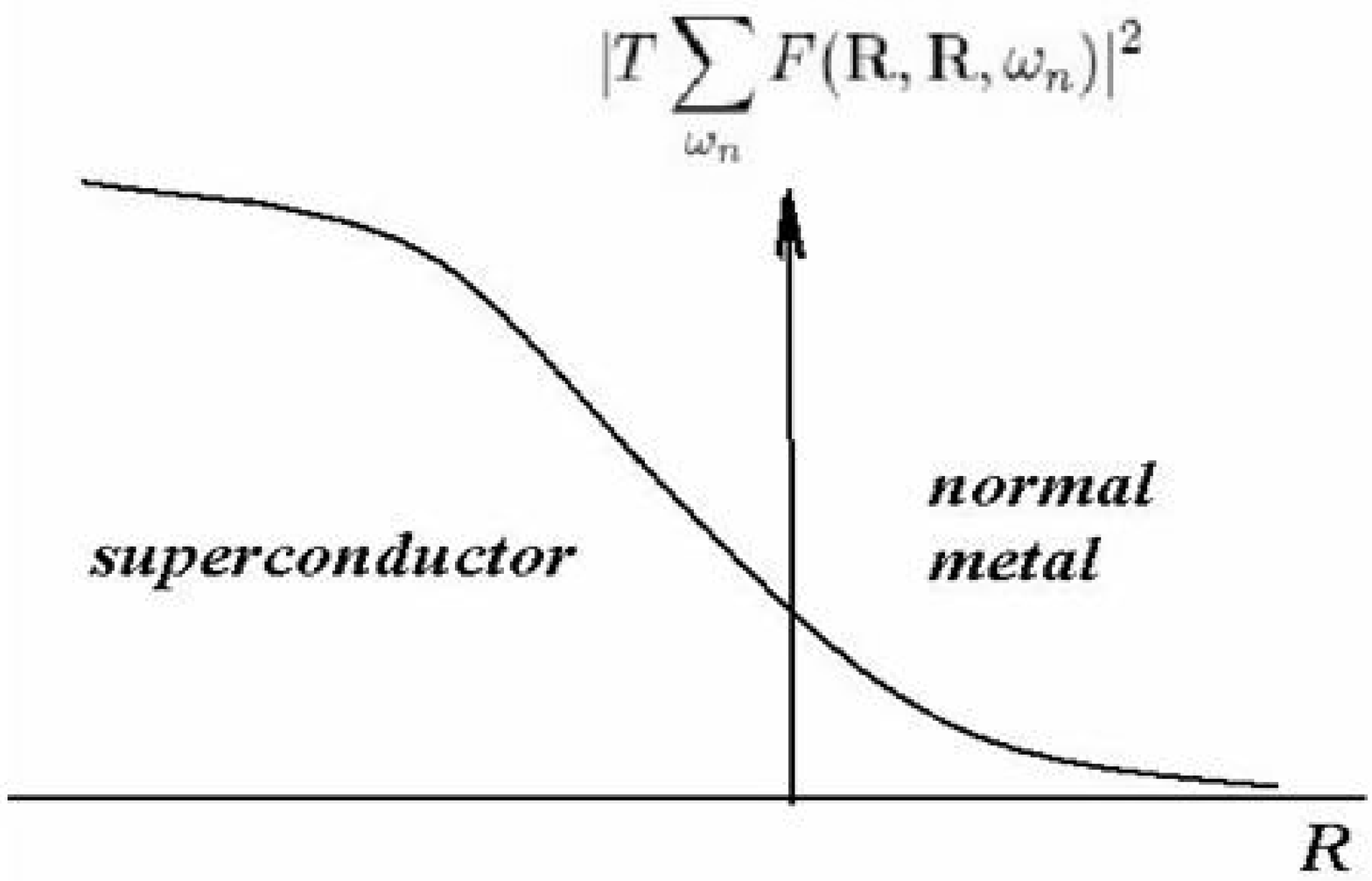} \includegraphics [width = 4cm] {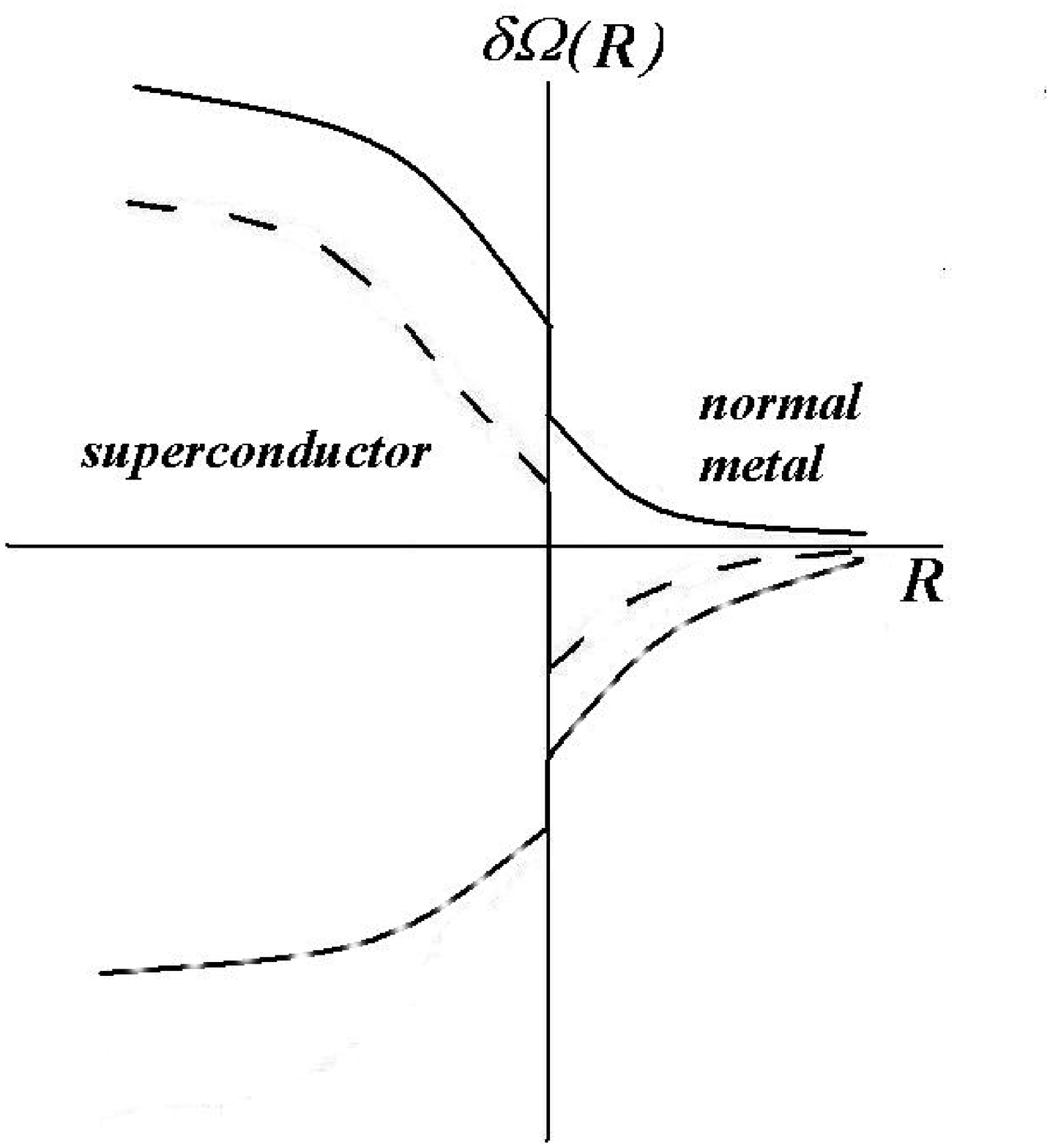}
\caption{(Left) schematic picture of Cooper pair density near the superconductor - normal metal interface. (Right) schematic picture of energy profile for impurities with different values of BCS interaction. 
Solid line corresponds to impurity with repulsive BCS interaction. The case of small attraction is shown by long dashed line. Short dashed line describes the case of strong attraction.}
\label{fig2}
\end{figure}

Taking this condition into account, we consider three main cases of the BCS constant at impurity, namely,
repulsion, weak and strong (with respect to the bulk value) attraction. The corresponding energy profiles are schematically shown in the right panel of Fig. \ref{fig2}.
The system with the local BCS repulsion impurities lowers the energy by expelling them from the superconducting region as shown by the solid line in \ref{fig2}.
Impurity with relatively weak BCS attraction expels from the region with strong
superconductivity into region of proximity-induced
superconductivity. Near the boundary there is a local minimum of the energy. The
case is shown by the long dashed line in Fig. \ref{fig2}.
Finally, it is energetically favourable for the impurity with strong BCS attraction to move into the bulk of a superconductor.
This case is shown by the short dashed line in Fig. \ref{fig2}.

Mobile impurity might have magnetic moment. This determines another mechanism
of energy profile formation near the superconductor-normal metal interface.
The exchange interaction of the magnetic impurity reads
\begin{equation}
H_{\mathrm{ex}}=J\mathbf{S}\cdot \boldsymbol{\sigma} (\mathbf{R}),
\end{equation}
where $\mathbf{S}$ is the spin operator of magnetic impurity, $\boldsymbol{\sigma }(
\mathbf{R})$ is the operator of the electronic spin density, and $J$ is the
exchange interaction constant, which is assumed to be isotropic and position independent.
The correction to the energy $\delta \Omega (R)\rightarrow \delta \Omega (R)+\delta \Omega_m (R)$ due to superconducting
correlations in the second order is given by
\begin{equation}
\delta \Omega_m (R)=J^{2}S(S+1)T\sum_{\omega _{n}}|F(\mathbf{R},\mathbf{R};\omega _{n})|^{2}  \label{mag-energy}.
\end{equation}
The sign of energy change in (\ref{mag-energy}) is positive, which means that magnetic
impurity in a superconductor has larger energy, than in normal metal in both cases of
ferromagnetic and antiferromagnetic exchange interaction constant. Physically, this is
related to the fact, that magnetic impurities destroy singlet
superconductivity.

At $T\sim \Delta (T)$ and distance $R \leq v_F/T$ we estimate for the
normal metal side of the interface
\begin{equation}\label{energy3}
\delta\Omega_m(R)=(J \nu)^{2}S(S+1)T \left[\Delta(T)/T \right]^{2}.
\end{equation}

The second order correction (\ref{mag-energy}) can be of the order of or greater than the correction due to the local BCS, since the latter (\ref{BCS-energy}) is counted from the average value.

Note, that we do not consider the case of very low temperatures, where the impurity mobility can be frozen.

In order to find Eq. (\ref{energy1},\ref{energy2},\ref{energy3}), we obtain the anomalous part of the Green function near the
superconductor-normal metal interface by solving the system of equations
\begin{equation}
\begin{bmatrix} i\omega_{n}+\frac{\boldsymbol{\nabla}^{2}}{2m}+\epsilon_{F} &
\Delta(\textbf{r},T)\\ \Delta^{*}(\textbf{r},T)&
i\omega_{n}-\frac{\boldsymbol{\nabla}^{2}}{2m}-\epsilon_{F} \end{bmatrix}G(\mathbf{r},\mathbf{r}';\omega_n)=\delta (\mathbf{r}-\mathbf{R}),
\end{equation}
where $\Delta (\mathbf{r},T)$ is a steplike function, which is zero in the region of
normal metal $x>0$ and constant $\Delta(T)$ in the superconductor $x<0$. The Cooper pair wave-function near superconductor-normal metal interface is shown schematically in the left panel of Fig. \ref{fig2}.

\section{Conclusion}
Let us now quickly comment on the energy profile at the vicinity of the superconducting vortex.
The vortex is characterized by a strong dependence of the superconducting gap function $\Delta(\mathbf{r},T)$ on the
distance to the center of the vortex and by the constancy of the interaction constant.
Therefore, the situations where impurities have local repulsion or strong attraction can be analyzed similarly as it was in case of superconductor-normal metal
interface.

We have studied the model of superconductor hosting mobile impurities, which locally modify the superconducting properties.
We find that the impurities that make up the mobile component attract to each other. Hence, there is a tendency to form a more condensed mobile component.
For impurities with local BCS attraction constant, which is smaller than that in the bulk of the superconductor, there is a minimum of energy near the superconductor - normal metal interface.
This increases the probability of finding such impurities near the interface.

We thank Alexander A. Zyuzin for interesting discussions.

\end{document}